\documentclass[10pt, conference]{IEEEtran}
\IEEEoverridecommandlockouts
\usepackage{amsmath,amssymb,amsfonts}
\usepackage{algorithmic}
\usepackage{graphicx}
\usepackage{textcomp}
\usepackage{xcolor}
\usepackage{booktabs}
\usepackage{balance}

\def\BibTeX{{\rm B\kern-.05em{\sc i\kern-.025em b}\kern-.08em
    T\kern-.1667em\lower.7ex\hbox{E}\kern-.125emX}}
\begin{document}

\title{NodeSRT: A Selective Regression Testing Tool for Node.js Application}

\author{\IEEEauthorblockN{Yufeng Chen}
\IEEEauthorblockA{
\textit{University of British Columbia}\\
Vancouver, Canada \\
yufengcy@student.ubc.ca}
}

\maketitle

\begin{abstract}
Node.js is one of the most popular frameworks for building web applications. As software systems 
mature, the cost of running their entire regression test suite can become significant. 
Selective Regression Testing (SRT) is a technique that executes only a subset of tests the regression test suite can detect software failures more efficiently. 
Previous SRT studies mainly focused on standard desktop applications. Node.js applications are 
considered hard to perform test reduction because of Node's asynchronous, event-driven programming model and because  
JavaScript is a dynamic programming language. 
In this paper, we present NodeSRT, a Selective Regression Testing framework for Node.js applications. 
By performing static and dynamic analysis, NodeSRT identifies the relationship between changed methods and tests, 
then reduces the regression test suite to only tests that are 
affected by the change to improve the execution time of the regression test suite. 
To evaluate our selection technique, we applied NodeSRT to two open-source projects: Uppy and Simorgh, 
then compared our approach with the retest-all strategy and current industry-standard SRT technique: Jest 
OnlyChange. The results demonstrate that NodeSRT correctly selects affected tests based on 
changes and is 250\% faster, 450\% more precise than the Jest OnlyChange. 
    
\end{abstract}

\begin{IEEEkeywords}
JavaScript, Selective Regression Testing, Node.js Application, Static Analysis, Dynamic Analysis
\end{IEEEkeywords}

\section{Introduction}
With the continuous growth of web applications, Node.js has become one of the most popular frameworks 
for web application development \cite{b16}. For critical online services, performing regression testing and integration testing is important. However, 
since JavaScript is a loosely typed, dynamic language, test 
selection on JavaScript projects is hard. Besides, modern web applications are usually composed of 
more than one component; running unit tests only does not judge the overall behaviour of the web 
application \cite{b8}. 

There are two phases involved in SRT. The first phase is to select tests based on the 
test dependency graph and changes. The second phase is to run selected tests.
Test selection techniques operate at four levels of granularity: statement, method, file, module. The common 
two are method-level and file-level. File-level granularity analysis builds a relationship between tests and files in 
the system and selects tests that reflect changed files. The method-level analysis builds a relationship between tests and methods 
then selects tests that are affected by changed methods. Since method-level selection is more complicated than  
file-level selection, file-level selection runs faster in phase one. However, file-level selection selects more 
tests than needed. Therefore it is less precise than method-level selection and runs slower in phase two \cite{b6}.
Jest OnlyChange is the current industry-standard SRT technique that operates at file-level granularity. And it can reduce tests executed 
without skipping tests that might expose failures. Plus it is the most light-weighted approach. Although fast, this approach may not be precise 
enough for some test suites. Therefore, our research starts from a question:\textit{``Can we find a more 
effective test selection technique for Node.js Applications?"} 

To evaluate the effectiveness of SRT techniques, Rothermel proposed four metrics: Inclusiveness, Precision, Efficiency, Generality \cite{b13}. Inclusiveness 
measures the extent to which SRT technique chooses tests that are affected by the change. Precision measures 
the ability that the SRT technique omits tests that are not affected by the change. Efficiency measures the time and 
space required. Generality measures its ability to function in a comprehensive and practical range of situations.
We say a selection technique is safe if it achieves 100\% inclusiveness.

Our intuition for reducing the total running time is to 
improve the granularity of the selection technique to improve precision so that fewer tests are required to run when the regression suite is executed. 
We also evaluated our selection technique by performing an empirical study on two open-source Node.js projects with different 
sizes and code coverage.

\section{Approach Overview}
To mitigate the challenge of performing test selection on JavaScript programs, our tool uses a combination of static and dynamic analysis, then performs a modification-based test 
selection algorithm at method level. The modification-based approach works by analyzing modified code entities 
to select tests based on modifications. This strategy can guarantee safety while being relatively simple. NodeSRT consists of five parts: dynamic analysis, static analysis, change analysis, 
test selector, and selected test runner. 
\begin{figure}[htbp]
    \centerline{\includegraphics[scale=0.45]{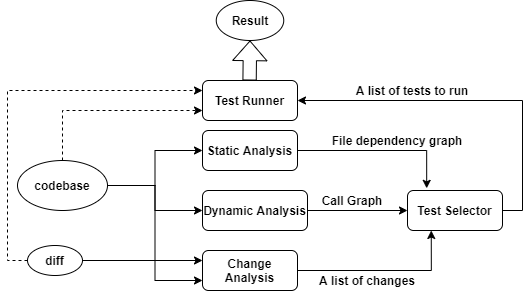}}
    \caption{NodeSRT Architecture}
    \label{fig}
    \end{figure}    

 The \textbf{Static Analysis module} performs static analysis on the original codebase to generate file dependency
graph on each test by identifying and resolving \verb|require| and \verb|import| in JavaScript files. 
The \textbf{Dynamic Analysis module} generates a dynamic call graph by injecting code to the original  
generated AST. 
NodeSRT uses HTTP requests to collect runtime information of the application. Since web applications usually consist of different modules, code in 
different modules may be running in different runtime environments, for example, the server-side code runs in the Node.js 
environment. Client-side code runs in the browser environment. The code injector in the Dynamic Analysis module injects code that sends logging messages to 
the logging server, which collects all logging messages and generates call graph in JSON format. The runtime information we collected includes the function name, file name, and the number of parameters. 
These entities are used to create a dynamic call graph. When the codebase becomes large, code analysis result should be store in a database to ensure performance \cite{b2, b4}. 
The \textbf{Change Analysis module} compares the ASTs of the changed files, then generate a list of changes in JSON format. Since NodeSRT uses function-level 
granularity, change analysis module finds the closest function name of each different AST node based on their ancestors. 
If the function is anonymous, NodeSRT will generate a unique name for it based on its parent function name, class name, and file name. 
This approach is similar to the approach for Chianti handling anonymous class in Java \cite{b12}. With call graph, file dependency graph, 
and JSON representation of changes, The \textbf{Test Selector} selects tests based on the list of changes and the call graph. To handle changes outside 
functions, the test selector selects tests that depend on the changed files based on file dependency graph to guarantee safety. 
Finally, \textbf{Test Runner} runs selected tests.

Our tool can also be used to select end-to-end tests since NodeSRT uses HTTP requests 
record runtime information and build dynamic call graph.

\section{Empirical Evaluation}

To evaluate NodeSRT, we performed an empirical evaluation on two open-source Node.js projects. We chose these two projects because 
our empirical study requires systems that have to be well-maintained and have reasonable amount of tests. By using the method mentioned in \cite{b10}, 
we selected Uppy and Simorgh. Uppy has 112k lines of code, 216 unit tests, and 9 end-to-end tests, 
achieves 20\% of code coverage. Simorgh has 698k lines of code. 
It includes 2801 unit tests, achieves 97\% of code coverage. The 
experiment ran on a 4 core x86-64 CPU with 16 GB of RAM, AWS cloud Linux server. Due to the fact that the internet speed and computing speed is not 
unchanged, we use the percentage of tests selected and the percentage of SRT full process running time to 
represent the result. 

We performed test selection on a total of 588 commits from the two subjects. For each commit, we generate a diff patch from the previous commit to serving as input to NodeSRT. 
Table I compares NodeSRT and Jest on average for selected tests and total running time. As we can see, given file dependency graph and call graph, the selection step 
for both projects is less than 5\% of total running time. Comparing to Jest OnlyChange, NodeSRT
selects much fewer tests for both projects. NodeSRT selects 1.5 times fewer 
tests in Uppy, 5.3 times fewer tests in Simorgh. Although NodeSRT selects less tests in Uppy, Jest OnlyChange runs faster than NodeSRT. 
This is because Jest OnlyChange makes use of Jest's own jest-haste-map module and customized file system module: watchman. Future works can 
be done for NodeSRT in this part. For project with high code coverage: Simorgh, NodeSRT selected fewer tests and is 2.7 times faster. 
\begin{table}[htbp]
    \caption{Empirical study result}
    \centering
    \begin{tabular}{@{}p{2em}p{2.2em}p{2em}p{1.8em}p{3.5em}p{2em}p{3.5em}p{3em}@{}}
    \toprule
    \textbf{Project Name} & \textbf{Round of exp.}  & \textbf{Retest time (s)} & \textbf{Select time (ms)} & \textbf{NodeSRT test No. (\%)}    & \textbf{Jest test No. (\%)}       & \textbf{NodeSRT running time (\%)} & \textbf{Jest run time (\%)} \\ \midrule
    Uppy         & \multicolumn{1}{r}{480} & \multicolumn{1}{r}{69.3}  & \multicolumn{1}{r}{490} & \multicolumn{1}{r}{8.4} & \multicolumn{1}{r}{20.7} & \multicolumn{1}{r}{50.2}  &   \multicolumn{1}{r}{23.6} \\ 
    Simorgh      & \multicolumn{1}{r}{108} & \multicolumn{1}{r}{450.7} & \multicolumn{1}{r}{2256} & \multicolumn{1}{r}{2.7}& \multicolumn{1}{r}{17.2} & \multicolumn{1}{r}{8.1}   &  \multicolumn{1}{r}{30.2} \\ \bottomrule
    \end{tabular}
    \end{table}

\section{Related Work and Conclusion}
There are several techniques proposed for standard desktop applications. These techniques 
first classify programs into different entities such as functions, types, variables, and macros, then utilize comprehensive static analysis 
and dynamic analysis to build entity-tests relationships to reduce test suite (e.g., \cite{b13, b2, b4,b12,b7, b9,b14, b17}).
For studies focusing on JavaScript applications, Mutandis is a generic mutation testing 
approach for JavaScript that guides the mutation generation process \cite{b11}. It works by leveraging static and dynamic program 
analysis to guide the mutation generation process a-priori towards parts of the code that are error-prone or 
likely to influence the program’s output. 
Tochal is a DOM-Sensitive change impact analysis tool for JavaScript. Through dynamic code injection and static analysis, it 
incorporates a ranking algorithm for indicating the importance of each entity in the impact set. This 
approach focused on frontend DOM changes rather than the frontend backend interaction \cite{b1}.

\textbf{Conclusion.} We present NodeSRT, a novel approach for performing SRT on Node.js applications at method level. Using a change-based selection technique, 
obtaining a function call relationship with dynamic analysis, collecting file dependency with static analysis, NodeSRT reduces regression tests in short running time
and high inclusiveness and precision. Empirical evaluation showed that our approach outperformed Jest OnlyChange in precision and total running time. Future work can 
be done in integrating our technique with unit testing frameworks to improve its performance further.

\balance
\bibliographystyle{IEEEtran}
\bibliography{reference}{}

\begin{thebibliography}{10}
\providecommand{\url}[1]{#1}
\csname url@samestyle\endcsname
\providecommand{\newblock}{\relax}
\providecommand{\bibinfo}[2]{#2}
\providecommand{\BIBentrySTDinterwordspacing}{\spaceskip=0pt\relax}
\providecommand{\BIBentryALTinterwordstretchfactor}{4}
\providecommand{\BIBentryALTinterwordspacing}{\spaceskip=\fontdimen2\font plus
\BIBentryALTinterwordstretchfactor\fontdimen3\font minus
  \fontdimen4\font\relax}
\providecommand{\BIBforeignlanguage}[2]{{%
\expandafter\ifx\csname l@#1\endcsname\relax
\typeout{** WARNING: IEEEtran.bst: No hyphenation pattern has been}%
\typeout{** loaded for the language `#1'. Using the pattern for}%
\typeout{** the default language instead.}%
\else
\language=\csname l@#1\endcsname
\fi
#2}}
\providecommand{\BIBdecl}{\relax}
\BIBdecl

\bibitem{b16}
F.~Schiavio, H.~Sun, D.~Bonetta, A.~Ros\`{a}, and W.~Binder, ``Nodemop: Runtime
  verification for node.js applications,'' in \emph{Proceedings of the
  Symposium on Applied Computing (SIGAPP)}, 2019, p. 1794–1801.

\bibitem{b8}
M.~Hirzel, ``Selective regression testing for web applications created with
  {G}oogle {W}eb {T}oolkit,'' in \emph{Proceedings of the International
  Conference on Principles and Practices of Programming on the Java Platform:
  Virtual Machines, Languages, and Tools (PPPJ)}, 2014, p. 110–121.

\bibitem{b6}
M.~Gligoric, L.~Eloussi, and D.~Marinov, ``Practical regression test selection
  with dynamic file dependencies,'' in \emph{Proceedings of the International
  Symposium on Software Testing and Analysis (ISSTA)}, 2015, p. 211–222.

\bibitem{b13}
G.~{Rothermel} and M.~J. {Harrold}, ``Analyzing regression test selection
  techniques,'' \emph{IEEE Transactions on Software Engineering (TSE)},
  vol.~22, no.~8, pp. 529--551, 1996.

\bibitem{b2}
A.~{Beszédes}, T.~{Gergely}, L.~{Schrettner}, J.~{Jász}, L.~{Langó}, and
  T.~{Gyimóthy}, ``Code coverage-based regression test selection and
  prioritization in webkit,'' in \emph{Proceedings of the International
  Conference on Software Maintenance (ICSM)}, 2012, pp. 46--55.

\bibitem{b4}
{Yih-Farn Chen}, D.~S. {Rosenblum}, and {Kiem-Phong Vo}, ``{TESTTUBE: A} system
  for selective regression testing,'' in \emph{Proceedings of International
  Conference on Software Engineering (ICSE)}, 1994, pp. 211--220.

\bibitem{b12}
{Xiaoxia Ren}, B.~G. {Ryder}, M.~{Stoerzer}, and F.~{Tip}, ``Chianti: {A}
  change impact analysis tool for {J}ava programs,'' in \emph{Proceedings
  International Conference on Software Engineering (ICSE)}, 2005, pp. 664--665.

\bibitem{b10}
A.~Labuschagne, L.~Inozemtseva, and R.~Holmes, ``Measuring the cost of
  regression testing in practice: {A} study of {J}ava projects using continuous
  integration,'' in \emph{Proceedings of the Joint Meeting on Foundations of
  Software Engineering (ESEC/FSE)}, 2017, p. 821–830.

\bibitem{b7}
M.~J. Harrold, J.~A. Jones, T.~Li, D.~Liang, A.~Orso, M.~Pennings, S.~Sinha,
  S.~A. Spoon, and A.~Gujarathi, ``Regression test selection for java
  software,'' \emph{SIGPLAN Not.}, vol.~36, no.~11, p. 312–326, Oct. 2001.

\bibitem{b9}
M.~J. {Harrold} and M.~L. {Souffa}, ``An incremental approach to unit testing
  during maintenance,'' in \emph{Proceedings of the Conference on Software
  Maintenance (ICSM)}, 1988, pp. 362--367.

\bibitem{b14}
G.~Rothermel and M.~J. Harrold, ``A safe, efficient regression test selection
  technique,'' \emph{ACM Transactions on Software Engineering Methodology
  (TOSEM)}, vol.~6, no.~2, p. 173–210, Apr. 1997.

\bibitem{b17}
A.~{Taha}, S.~M. {Thebaut}, and S.~{Liu}, ``An approach to software fault
  localization and revalidation based on incremental data flow analysis,'' in
  \emph{Proceedings of the International Computer Software Applications
  Conference (SAC)}, 1989, pp. 527--534.

\bibitem{b11}
S.~{Mirshokraie}, A.~{Mesbah}, and K.~{Pattabiraman}, ``Efficient {JavaScript}
  mutation testing,'' in \emph{Proceedings of the International Conference on
  Software Testing, Verification and Validation (ICST)}, 2013, pp. 74--83.

\bibitem{b1}
S.~Alimadadi, A.~Mesbah, and K.~Pattabiraman, ``{Hybrid DOM-Sensitive Change
  Impact Analysis for JavaScript},'' in \emph{European Conference on
  Object-Oriented Programming (ECOOP)}, vol.~37, 2015, pp. 321--345.

\end{thebibliography}

\vspace{12pt}

\end{document}